\documentclass[prc,twocolumn,showpacs,nofootinbib]{revtex4}
\usepackage{amsmath}
\usepackage{amssymb}
\usepackage{braket}
\usepackage{color}

\usepackage{graphicx}
\mathchardef\up"0222
\mathchardef\down"0223
\begin{document}
\bibliographystyle{apsrev}
 
\title{Spin response and neutrino mean free path in neutron matter}

\author{Luca Riz}
\email{luca.riz@unitn.it}
\affiliation{Dipartimento di Fisica, University of Trento, via Sommarive 14, 
I--38123,
Povo, Trento, Italy}
\affiliation{INFN-TIFPA, Trento Institute for Fundamental Physics and Applications, Trento, Italy}
 
\author{Stefano Gandolfi}
\email{stefano@lanl.gov}\affiliation{Theoretical Division, Los Alamos National Laboratory, Los Alamos, NM}

\author{Francesco Pederiva}
\email{francesco.pederiva@unitn.it}\affiliation{Dipartimento di Fisica, University of Trento, via Sommarive 14, 
I--38123,
Povo, Trento, Italy}
\affiliation{INFN-TIFPA, Trento Institute for Fundamental Physics and Applications, Trento, Italy}

\date{\today}

\begin{abstract} 
The longitudinal and transverse density- and spin-density response functions in Pure Neutron Matter (PNM) are derived over a wide range of densities within the Time Dependent Local Spin Density Approximation (TDLSDA). The underlying density functional was derived from an Auxiliary Field Diffusion Monte Carlo (AFDMC) computation of the equation of state of unpolarized and fully spin polarized pure neutron matter. In order to assess the dependence of the results on the specific underlying nucleon-nucleon Hamiltonian, we used both the phenomenological Argonne AV8$^{\prime}$+UIX force, and local chiral forces up to N$^2$LO. The resulting response function has then been applied to the study of the neutrino mean free path in PNM. 
\end{abstract}


\maketitle

\section{Introduction}
As shown many years ago, the Weinberg-Salam Lagrangian \cite{Wei72} describing the interaction of neutrinos with baryonic matter can be translated, after a non relativistic reduction, into the calculation of response to density and spin/isospin density operators \cite{Saw75}. Several non relativistic many-body calculations have been carried out over time \cite{Bur98}, in particular via a direct evaluation of the propagator in the context of the use of Skyrme-like forces \cite{Pastore1,Pastore2,Pastore3}, or by extending the Tamm-Dancoff approximation to the inclusion of dynamical correlations \cite{Cow03,Ben07,Lov13}. At present, computing response functions in a many-body system within an {\it ab-initio} scheme is technically possible, but still quite expansive from the computational point of view. In a previous paper \cite{Lip13} a fair compromise was devised between including the whole microphysics, that is usually addressable in ground state calculations, and a purely mean field treatment. This was achieved following the standard prescription suggested by the Hohenberg-Kohn theorem to obtain a realistic, though simplified, density functional, and employing it within the Time Dependent Local Density Approximation (TDLDA). The first application was the study of the contribution of the longitudinal \cite{Lip13} isospin channel to the neutrino cross section in nuclear matter with an arbitrary value of the isospin asymmetry parameter. The TDLDA approximation has been also applied to the study of the transverse \cite{Lip16} isospin channel response functions for an arbitrary isospin asymmetry parameter.

In this paper we extend the TDLDA approach to the study of the density and spin-density response functions in pure neutron matter (PNM), both in the longitudinal and in the transverse channels. Also in this case we do not limit our study to the unpolarized and fully-polarized cases, but we consider arbitrary spin polarization. This is an extension of a formalism that is very well known, particularly in condensed matter applications, known as Time Dependent Local Spin Density Approximation (TDLSDA)(see e.g. \cite{Lip03}). 
The key ingredient of any mean field calculation based on the Local Density Approximation (LDA) is the determination of an accurate density functional based on a pre-existing Equation of State (EoS) $E[\rho]$. Following the Hohenberg-Kohn prescription, $E[\rho]$ can be extracted for a homogeneous system by simply fitting the exact energy as a function of the (spin/isospin-)density. By subtracting the energy of the Free Fermi Gas (FFG) at the same density it is then possible to obtain the non-trivial part of the energy density functional. The Local Density Approximation (LDA) allows then to address problems for inhomogeneous systems and excited states. 

In this paper we also want to check the robustness of the TDLSDA predictions against the underlying functional. In order to do that, we use two different Hamiltonians. The first includes a phenomenological two- plus three-neutron interaction (namely AV8'+UIX). The second employs modern local chiral EFT potentials up to N$^2$LO \cite{Gez13, Gez14, Lyn16, Tew16, Tew18}, and estimating the systematic error due to the uncertainty on the model potential. 

The paper is organized as follows. In Sec. II we describe in some details the procedures and the results concerning the computation of the EoS for the Hamiltonians considered. In Sec. III We briefly revise the formalism for computing the TDLSDA response function in both the longitudinal and transverse channels.  Sec. IV shows the numerical results for both the response functions and the neutrino mean free path. Sec. V is devoted to conclusions.

\section{Equation of State}

The first step in this analysis is the computation of the Equation of State. This is achieved by means of Auxiliary Field Diffusion Monte Carlo Methods \cite{Sch99,Gan09}. As previously mentioned, two different nucleon-nucleon interaction schemes have been used. The first EoS (EOSA thereafter) is derived from the well known Argonne AV8$^{\prime}$ potential for the two-body interaction, plus the Urbana UIX interaction for the three body channel. This interaction has been widely used to study homogeneous neutron matter and nuclear matter properties (see \cite{Gan15,Car15} and references therein). The second EoS (EOS$\chi$ thereafter) is based on potentials derived within Chiral Effective Field theory ($\chi-EFT$).
Among different implementations of the Effective Chiral potential which have been recently developed, we chose a local formulation up to N$^2$LO which have been derived by A. Gezerlis et al. \cite{Gez13,Gez14}.

\begin{figure}[hbt!]

\centerline{\includegraphics[scale=0.34]{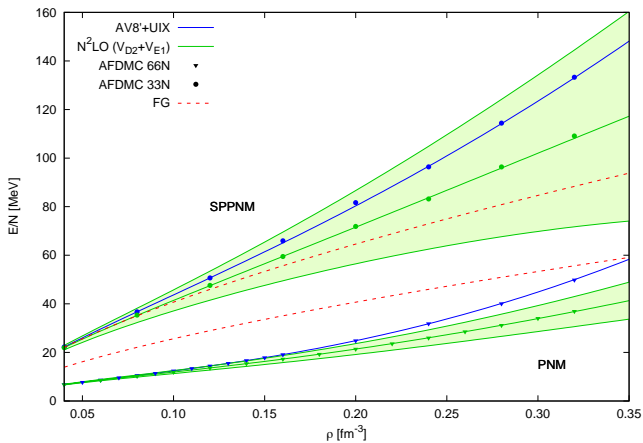}}

\caption[]{Equation of State for PNM (lower set) and for SPPNM (upper set) with AV8$^{\prime}$+UIX (blue curves) and with Chiral Potentials at N$^2$LO (green bands). More details on the potentials are described in the text. Errorbars for the Chiral effective interaction have been computed according to Epelbaum et al. \cite{Epe15}. For the sake of comparison, we also report the corresponding curves for a free Fermi gas at the thermodynamic limit (red dotted curves).
}
\label{fig1}
\end{figure}

In Fig. \ref{fig1} we report the results obtained from our calculations of a fully Spin Polarized Pure Neutron Matter (SPPNM) for densities ranging from $0.04$ fm$^{-3}$ up to $2\rho_0$, where $\rho_0=0.16$ fm$^{-3}$ is the nuclear saturation density.  The results for Pure Neutron Matter (PNM) are those obtained by Gandolfi et al. \cite{Gan14} and I. Tews et al. \cite{Tew16} for the phenomenological and the chiral interaction respectively. For the chiral potential we used the N$^2$LO(D2,E1) cutoff $R_0=1.0$ fm, $c_e=0.62$ $c_d=0.5$ as described in \cite{Lyn16}. SPPNM energies were computed for A=33 neutrons in a periodic box. In order to reduce the impact of finite size effects, the potential was computed by a sum over the first neighbors of a given simulation cell. The statistical errors of the data reported in Fig.1 are of the size of the symbols. The bands relative to the chiral potential results have been obtained using the prescription of Epelbaum et al. \cite{Epe15}. The Equation of state representing the upper and lower limits of the band are denoted as EOS$\chi_u$ and EOS$\chi_l$ respectively.
The errorbars are larger in the spin polarized EOS since the estimated theoretical error scales like $k_F^3$ up to next-to-next-to-leading order and at fixed density the Fermi sphere has to be filled up to larger values of momentum for polarized systems.
The EoS we computed for the polarized system is reasonable compared to the one obtained by Kr\"{u}ger at al. \cite{Kru15}, at least up to saturation density.
In our discussion we will only consider the density range $0.5\rho_0\le\rho\le2\rho_0$. Another interesting feature to be noticed in the comparison is that the spin symmetry energy, given by the difference between the energy per neutron of the spin polarized and spin unpolarized systems tends to be substantially larger in EOSA than in EOS$\chi$.

The Monte Carlo results are fitted in order to derive the energy density functional to be used in the TDLSDA response function. We recall that in the Local Density (mean field) approximation, the energy as a function of the density $\rho$ and the spin polarization $\xi$ can be generically written as: 
\begin{equation}
E(\rho,\xi)=T_0(\rho,\xi)+\int\epsilon_V(\rho,\xi)\rho d {\bf r}.
\label{DF}
\end{equation}
The quantities $\rho$ and $\xi$ are related to the density of particles with spin up $\rho_\uparrow$ and the density of particle with spin down $\rho_\downarrow$ in the following way:
\begin{eqnarray}
\rho = \rho_\uparrow+\rho_\downarrow;\nonumber\\
\\
\xi = \frac{\rho_\uparrow-\rho_\downarrow}{\rho}\nonumber.
\end{eqnarray}

\noindent
We define the functional $\epsilon(\rho,\xi)$ using the common assumption of a quadratic dependence on the spin polarization:
\begin{equation}
\epsilon_V(\rho,\xi)=\epsilon_0(\rho)+\xi^2\left[\epsilon_1(\rho)-\epsilon_0(\rho)\right],
\end{equation}
where the functions $\epsilon_i$ are defined as polynomials in the neutron density:
\begin{eqnarray}
\begin{array}{cl}
\displaystyle
\epsilon_i(\rho)=&\epsilon^0_i+a_i\left(\frac{\rho-\rho_0}{\rho_0}\right)+b_i\left(\frac{\rho-\rho_0}{\rho_0}\right)^2+c_i\left(\frac{\rho-\rho_0}{\rho_0}\right)^3 \label{eps_q}
\end{array}
\end{eqnarray}
Such functions will contain the whole information about the interaction, i.e. all the terms that in ordinary LDA theory are separately referred to as "direct", "exchange", and "correlation" terms.
The index $i=0,1$ will indicate the spin unpolarized and polarized neutron matter ($\xi=0,1$) respectively.
As usual, we assume the value of the saturation density to be $\rho_0=0.16$ fm$^{-3}$.
Despite there is no implicit or explicit expectation of a hierarchical ordering in our expansion of the density functional, the coefficients fitted on the numerical AFDMC results for EOSA and EOS$\chi$, reported in Tab. \ref{fits}, show some prevalence of the first and second order expansion terms (apart for EOS$\chi_l$ for SPPNM, which has $b_i$ and $c_i$ of the same order).

\begin{table}[h!]
    \begin{tabular}{llrrrr}
    \hline\hline
    EOSA &       & $\epsilon^0_i$ & $a_i$ & $b_i$ & $c_i$ \\\hline
    (SPPNM) & $i$=1   & 9.411 & 21.997 & 13.032 & 0.262 \\
    (PNM) & $i$=0   & -15.97 & -2.689 & 12.435 & 0.521 \\
\hline\hline
    EOS$\chi$   &       & $\epsilon^0_i$ & $a_i$ & $b_i$ & $c_i$ \\\hline
    (SPPNM) & $i$=1   & 3.85  & 10.975 & 6.433 & -1.506 \\
    (PNM) & $i$=0   & -17.81 & -7.865 & 7.746 & -0.934 \\ \hline\hline    
    EOS$\chi_u$  &       & $\epsilon_0$ & $a_i$ & $b_i$ & $c_i$ \\\hline
    (SPPNM) & $i$=1   & 13.87  & 29.206 & 14.573 & -1.502 \\
    (PNM) & $i$=0   & -16.36 & -5.117 & 9.367 & -0.565 \\\hline\hline
    EOS$\chi_l$  &       & $\epsilon^0_i$ & $a_i$ & $b_i$ & $c_i$ \\\hline
    (SPPNM) & $i$=1   & -6.18 & -7.256 & -1.707 & -1.510 \\
    (PNM) & $i$=0   & -19.26 & -10.614 & 6.126 & -1.303 \\\hline\hline
    \end{tabular}%
  \label{tab:addlabel}%
  \caption{Coefficient fitting the density functional of Eq.(\ref{eps_q}) to the
  EoS computed by means of the AFDMC method. EOSA refers to the EoS from the AV8'+UIX potential, while the EOS$\chi$ are relative to the Hamiltonian with the local N$^2$LO(D2,E1) chiral interaction. The three tables refer to the center, upper limit and lower limit of the uncertainty band respectively.}
  \label{fits}
\end{table}%

\section{Time Dependent Local Spin Density Approximation}

The density functional of Eq. {\ref{DF}} can be used to describe the density and spin density excitations by means of the Time Dependent Local Spin Density Approximation (TDLSDA). In the spirit of the mean field theory, the solution of the many-body Schroedinger equation for $N$ neutrons in a volume $V$ such that $\rho=N/V$ is assumed to be the product of two Slater determinants, one for the $N_\uparrow$ spin-up neutrons and one for the $N_\downarrow$ spin-down neutrons:
\begin{equation}
\Psi({\bf r}_1\dots{\bf r}_N) = {\rm det}[\phi^\uparrow_i({\bf r}_j)]{\rm det}[\phi^\downarrow_i({\bf r}_j)],
\end{equation}
where the indices $i,j$ run from 1 to $N^\uparrow$ and $N^\downarrow$ respectively.
The spin-up and spin-down neutron densities are defined as:
\begin{equation}
\rho_{\sigma}=\sum_i\left|\varphi_i^{\sigma}({\bf r})\right|^2,
\end{equation}
where $\sigma=\up(\down)$ stands for spin-up and spin-down neutrons, respectively, and the sum runs over all the occupied states.
By minimizing the energy functional of Eq.(\ref{DF}) with respect to the single particle wavefunction $\varphi^\sigma_i$, one obtains 
the set of self-consistent, stationary Kohn-Sham equations for spin-up and spin-down neutrons wave functions ($\hbar=c=1$ hereafter):
\begin{equation}
\left[-\frac{1}{2m}\nabla^2_{\bf r}+v({\bf r})+w({\bf r})\sigma_z+\frac{1}{2}\omega_L\sigma_z\right]\varphi_i^{\sigma}({\bf r})=\epsilon_{i,\sigma}\varphi_i^{\sigma}({\bf r}).
\label{Schroed_gen}
\end{equation}
The term containing $\omega_L$ is needed to induce a partial (or total) magnetization of neutrons, mimicking the presence of an external (magnetic) field. The effective potentials are defined as the derivatives of the functional with respect to the total density and the magnetization:
\begin{equation}
\label{eq 1}
v({\bf r})=\frac{\partial\rho\epsilon_V\left[\rho({\bf r}),\xi\right]}{\partial\rho({\bf r})},
\quad
w({\bf r})=\frac{\partial\epsilon_V\left[\rho({\bf r}),\xi\right]}{\partial\xi({\bf r})}.
\end{equation}
We will briefly review the derivation of the TDLSDA in the longitudinal and in the transverse channels. 
\subsection{Longitudinal channel.}
The longitudinal channel describes the response to a time-dependent field along the ${\bf r}$ direction:
\begin{equation}
F^z = \sum_{k=1}^{N}f({\bf r}_k)\lambda_{\sigma}^k,
\label{Excitation}
\end{equation}
where:
\[
f({\bf r})=\exp\left[i({\bf q}\cdot{\bf r}-\omega t)\right]+\exp\left[-i({\bf q}\cdot{\bf r}-\omega t)\right],
\]
and $\lambda_{\sigma}^k=\lambda$ for a {\it density excitation} and $\lambda_{\sigma}^k=\lambda\eta_{\sigma}$, $\eta_{\sigma}$ is the eigenvalue of the $\sigma_z$ operator ($\eta=1$ for spin-up and $\eta=-1$ for spin down neutrons) for {\it vector-density excitations}, $q$ is the momentum and $\omega$ is the energy.
The corresponding time dependent Kohn-Sham equations reads:
\begin{equation}
\begin{split}
i\frac{\partial}{\partial t}\varphi_i^{\sigma}({\bf r,t})=&
\biggl\{-\frac{1}{2m}\nabla^2_{\bf r}+v\left[\rho_{\up}({\bf r},t),\rho_{\down}({\bf r},t)\right]\\
&+w\left[\rho_{\up}({\bf r},t),\rho_{\down}({\bf r},t)\right]\eta_{\sigma}\\
&+\lambda_{\sigma}\left[e^{i({\bf q}\cdot{\bf r}-\omega t)}+e^{-i({\bf q}\cdot{\bf r}-\omega t)}\right]\biggr\}\varphi_i^{\sigma}({\bf r},t).
\end{split}
\end{equation}
\noindent
For this case we use $\omega_L=0$, since longitudinal excitations are not directly coupled to the neutron spin. The solutions linearized in the neutron density oscillations induced by external fields are given by:
\begin{eqnarray}
\begin{array}{cl}
\rho_{\up}({\bf r},t)=&\rho_{\up}+\delta\rho_{\up}({\bf r},t),\\
\rho_{\down}({\bf r},t)=&\rho_{\down}+\delta\rho_{\down}({\bf r},t),
\end{array}
\end{eqnarray}
where the time dependent density is assumed to be proportional to the external perturbation: 
\begin{eqnarray}
\begin{array}{cl}
\delta\rho_{\up}({\bf r},t)=&\delta\rho_{\up}(e^{i({\bf q}\cdot{\bf r}-\omega t)}+e^{-i({\bf q}\cdot{\bf r}-\omega t)}),\\
\delta\rho_{\down}({\bf r},t)=&\delta\rho_{\down}(e^{i({\bf q}\cdot{\bf r}-\omega t)}+e^{-i({\bf q}\cdot{\bf r}-\omega t)}).
\end{array}
\end{eqnarray}


Following the derivation in Ref. \cite{Lip13}, the density-density response (per unit volume) is then given by:
\begin{equation}
\frac{\chi^s(q,\omega)}{V}=\frac{(\delta\rho_{\up}+\delta\rho_{\down})}{\lambda}\equiv\chi^{\up}(q,\omega)+\chi^{\down}(q,\omega),
\end{equation}
and the vector density-vector density response is:
\begin{equation}
\frac{\chi^v(q,\omega)}{V}=\frac{(\delta\rho_{\up}-\delta\rho_{\down})}{\lambda}\equiv\chi^{\up}(q,\omega)-\chi^{\down}(q,\omega).
\end{equation}
In order to determine the expression of the response function, we can explicitly compute the total self-consistent potentials in the Kohn-Sham equations. At first order in $\delta\rho_\sigma$ this is given by: 
\begin{equation}
\begin{split}
V_{KS}&\left[\rho_{\up}({\bf r},t),\rho_{\down}({\bf r},t)\right]\equiv v[\rho_\uparrow,\rho_\downarrow]+w[\rho_\uparrow,\rho_\downarrow]=\\
&=V_{KS}(\rho_{\up},\rho_{\down})+\left.\frac{\partial V_{KS}}{\partial\rho({\bf r},t)}\right|_{\rho_{\up},\rho_{\down}}\delta\rho_{\up}({\bf r},t)+\\
&+\left.\frac{\partial V_{KS}}{\partial\rho({\bf r},t)}\right|_{\rho_{\up},\rho_{\down}}\delta\rho_{\down}({\bf r},t),
\end{split}
\label{KSdyn_1}
\end{equation}
which gives the following expression for the Kohn-Sham equations:
\begin{eqnarray}
i\frac{\partial}{\partial t}\varphi_i^{\up}({\bf r},t)=&
\biggl\{-\frac{1}{2m}\nabla^2_{\bf r}+\textrm{const.}+\left[\delta\rho_{\up}V_{\up,\up}+\delta\rho_{\down}V_{\up,\down}+\lambda\right]\nonumber\\
& \times (e^{i({\bf q}\cdot{\bf r}-\omega t)}+e^{-i({\bf q}\cdot{\bf r}-\omega t)})\biggr\}\varphi_i^{\up}({\bf r},t),\nonumber\\\\
i\frac{\partial}{\partial t}\varphi_i^{\down}({\bf r},t)=&
\biggl\{-\frac{1}{2m}\nabla^2_{\bf r}+\textrm{const.}+\left[\delta\rho_{\up}V_{\up,\down}+\delta\rho_{\down}V_{\up,\up}\pm\lambda\right]\nonumber\\
& \times (e^{i({\bf q}\cdot{\bf r}-\omega t)}+e^{-i({\bf q}\cdot{\bf r}-\omega t)})\biggr\}\varphi_i^{\down}({\bf r},t),\nonumber
\label{KSdyn_2}
\end{eqnarray}
where the constant term is the Kohn-Sham potential evaluated at the density and magnetization of the homogeneous neutron matter under consideration. This fact makes the solutions of the linearized dynamic equations equal to those of the free Fermi gas. As a consequence, the response function of the system will be the one for the free system $\chi_0(q,\omega)=\chi_0^\uparrow(q,\omega)+\chi_0^\downarrow(q,\omega)$, where:
\begin{equation}
\begin{split}
\chi_0^{\up}(q,\omega)=\frac{V\delta\rho_{\up}}{\lambda_{\up}^{\prime}},\\
\chi_0^{\down}(q,\omega)=\frac{V\delta\rho_{\down}}{\lambda_{\down}^{\prime}}.
\end{split}
\end{equation}
The effective strength $\lambda'_\sigma$, defined as:
\begin{equation}
\begin{split}
\lambda_{\up}^{\prime}=\delta\rho_{\up}V_{\up,\up}+\delta\rho_{\up}V_{\up,\down}+\lambda,\\
\lambda_{\down}^{\prime}=\delta\rho_{\up}V_{\down,\up}+\delta\rho_{\up}V_{\down,\down}\pm\lambda
\end{split}
\end{equation}
include terms depending on the interaction. The mean field potentials $V_{\sigma,\sigma'}$ are obtained through the derivatives of $v+\eta_\sigma w$ with respect to $\rho_\sigma$: 
\begin{eqnarray}
V_{\up,\up}&=\left.\frac{\partial (v+w)}{\partial\rho_{\up}({\bf r},t)}\right|_{\rho_{\up},\rho_{\down}}=
\left.\left(\frac{\partial}{\partial\rho}+\frac{1}{\rho}\frac{\partial}{\partial\xi}\right)(v+w)\right|_{\rho,\xi},\nonumber\\
V_{\up,\down}&=\left.\frac{\partial (v+w)}{\partial\rho_{\down}({\bf r},t)}\right|_{\rho_{\up},\rho_{\down}}=
\left.\left(\frac{\partial}{\partial\rho}-\frac{1}{\rho}\frac{\partial}{\partial\xi}\right)(v+w)\right|_{\rho,\xi},\nonumber\\
V_{\down,\up}&=\left.\frac{\partial (v-w)}{\partial\rho_{\up}({\bf r},t)}\right|_{\rho_{\up},\rho_{\down}}=
\left.\left(\frac{\partial}{\partial\rho}+\frac{1}{\rho}\frac{\partial}{\partial\xi}\right)(v-w)\right|_{\rho,\xi},\nonumber\\
V_{\down,\down}&=\left.\frac{\partial (v-w)}{\partial\rho_{\down}({\bf r},t)}\right|_{\rho_{\up},\rho_{\down}}=
\left.\left(\frac{\partial}{\partial\rho}-\frac{1}{\rho}\frac{\partial}{\partial\xi}\right)(v-w)\right|_{\rho,\xi}.\nonumber
\end{eqnarray}
Comparing Eq.(\ref{KSdyn_1}) and Eq.(\ref{KSdyn_2}) we immediately see that:
\begin{equation}
\begin{split}
\lambda\chi^{\up}(q,\omega)=\lambda_{\up}^{\prime}\chi_0^{\up}(q,\omega)=L\delta\rho_{\up},\\
\lambda\chi^{\down}(q,\omega)=\lambda_{\down}^{\prime}\chi_0^{\down}(q,\omega)=L\delta\rho_{\down}.
\end{split}
\end{equation}
The solution of these equations, finally gives the TDLSDA response functions in the longitudinal channel:
\begin{widetext}
\begin{eqnarray}
\begin{array}{c}
\displaystyle
\chi^{s}(q,\omega)=V\frac{\chi^{\up}_0[V-(V_{{\down}{\down}}-V_{{\up}{\down}})\chi^{\down}_0]
+\chi^{\down}_0[V-(V_{{\up}{\up}}-V_{{\down}{\up}})\chi^{\up}_0]}
{(V-V_{{\down}{\down}}\chi^{\down}_0)(V-V_{{\up}{\up}}\chi^{\up}_0)
-V_{{\up}{\down}}\chi^{\up}_0V_{{\down}{\up}}\chi^{\down}_0~},
\\ \\
\displaystyle
\chi^{v}(q,\omega)=V\frac{\chi^{\up}_0[V-(V_{{\down}{\down}}+V_{{\up}{\down}})\chi^{\down}_0]
+\chi^{\down}_0[V-(V_{{\up}{\up}}+V_{{\down}{\up}})\chi^{\up}_0]}
{(V-V_{{\down}{\down}}\chi^{\down}_0)(V-V_{{\up}{\up}}\chi^{\up}_0)
-V_{{\up}{\down}}\chi^{\up}_0V_{{\down}{\up}}\chi^{\down}_0}~.
\end{array}
\end{eqnarray}
\end{widetext}

In the low-$q$, low-$\omega$ limits the free response functions $\chi_0^{\up}$ and $\chi_0^{\down}$ can be expressed as:
\begin{equation}
\chi_0^{\up,\down}({\bf q},\omega)=-V\nu^{\up,\down}\left[1+\frac{s}{2(1\pm\xi)^{1/3}}\ln{\frac{s-(1\pm\xi)^{1/3}}{s+(1\pm\xi)^{1/3}}}\right],
\label{resc_chi}
\end{equation}
where 
$\nu^{\up,\down}=mk_F^{\up,\down}/(2\pi^2)=mk_F(1\pm\xi)^{1/3}/(2\pi^2)$, $k_F=(3\pi^2\rho)^{1/3}$ and $s=\omega/(qv_F)$.
By defining:
\begin{equation}
\Omega^{\up,\down}=\left[1+\frac{s}{2(1\pm\xi)^{1/3}}\ln{\frac{s-(1\pm\xi)^{1/3}}{s+(1\pm\xi)^{1/3}}}\right], 
\label{resc_omega}
\end{equation}
we can rewrite the density-density and vector-density/vector-density response functions as:
\begin{widetext}
\begin{equation}
\frac{\chi^{s,v}}{Nm/(2k_F^2)}=-3\frac{(1+\xi)^{1/3}\Omega^{\up}\left[1+(G_{\down}\mp(\frac{1-\xi}{1+\xi})^{1/6}G_{\up\down})\Omega^{\down}\right]+(1-\xi)^{1/3}\Omega^{\down}\left[1+(G_{\up}\mp(\frac{1+\xi}{1-\xi})^{1/6}G_{\down\up})\Omega^{\up}\right]}
{(1+G_{\down}\Omega^{\down})(1+G_{\up}\Omega^{\up})-G^2_{\up,\down}\Omega^{\up}\Omega^{\down}},
\label{CHI_LONG}
\end{equation}
\end{widetext}
where $G_{\up}=\nu_{\up}V_{\up,\up}$, $G_{\down}=\nu_{\down}V_{\down,\down}$ and $G_{\up\down}=\sqrt{\nu_{\up}\nu_{\down}}V_{\up,\down}$.

The imaginary part of Eq. (\ref{CHI_LONG}) provides the strength of the single particle excitations:
\begin{equation}
S(q,\omega) = -\frac{1}{\pi}\chi^{s,v}(q,\omega).
\label{S_LONG}
\end{equation}
\subsection{Transverse channel}

The derivation of the response function in the transverse channel is similar to  that used in the longitudinal channel \cite{Lip13}. The excitation operator has the same structure as that of Eq. (\ref{Excitation}), but the constraint now is that $\Delta S_z=\pm 1$, thereby defining:
\begin{equation}
    F^{\pm}=\sum_k f({\bf r_k})\sigma^{\pm}_k.
\end{equation}
The parameter $\omega_L$ in Eq. (\ref{Schroed_gen}) can be related to the spin asymmetry of the system $\xi$ ($\xi=m/\rho=(N_{\up}-N_{\down})/N$) by imposing that the variation of the LSDA energy with respect to $\xi$ be zero \cite{Orl98,Lip03}:
\begin{eqnarray}
\label{eq 11111}
\int d{\bf r}(\rho_{\up}-\rho_{\down})=N_{\up}-N_{\down}=\omega_L\frac{\frac {3 N}{4\epsilon_F}}{ 1+{\frac{3\rho}{2\epsilon_F}}\frac{\partial w}{\partial m}}~,
\end{eqnarray}
where $\epsilon_F=k_F^2/2m$ is the Fermi energy, with the Fermi momentum $k_F$ and
the spin-up and spin-down neutron momenta given by $k_F^{\up}=k_F(1+\xi)^{1/3}$ and $k_F^{\down}=k_F(1-\xi)^{1/3}$, respectively. 

The derivation of the transverse response function was carried out first by
Rajagopal \cite{Raj78}, and was applied to quantum dots by Lipparini
et~al. \cite{Lip98,Lip99}. 

In the $\Delta S_z=\pm1$ channel, given the magnetization $m$ of the system,
the static LSDA equations can be rewritten as:
\begin{eqnarray}
\label{eq 3}
\left[\rule{0cm}{0.5cm}\right.
-\frac{1}{2} \nabla_{\bf r}^2  
+\frac{1}{2}\omega_L\sigma_z
+v({\bf r}) +{\cal W} {\bf m}\cdot {\bf \sigma}
\left.\rule{0cm}{0.5cm}\right]\varphi^{\sigma}_i({\bf r})
=\varepsilon_{i,\sigma}\,\varphi^{\sigma}_i({\bf r})\; ,
\end{eqnarray}
where ${\bf m}$ is the {\it spin polarization vector}. The interaction/correlation energy only depends on $\rho$ and
$|{\bf m}|$, i.e. $\epsilon_V=\epsilon_V[\rho, |{\bf m}|]$ so that the
isospin-dependent interaction/correlation potential $w$ in equation
(\ref{eq 1}) can be written as:
\begin{equation}
{\cal W} {\bf m}=w[\rho, |{\bf m}|]\, {\bf m}/|{\bf m}|\,,
\end{equation}
where:
\begin{equation}
w[\rho, |{\bf m}|]= \partial \epsilon_V[\rho,|{\bf m}|]\,/\partial|{\bf m}|\,,
\end{equation}
and ${\cal W}[\rho, |{\bf m}|] \equiv w[\rho,
|{\bf m}|]/|{\bf m}|$. 
Defining the spherical components $\pm$ of the
vectors ${\bf m}$ and ${\bf \sigma}$, it is possible to express the $z$ component of the magnetization dependent potential as:
\begin{eqnarray}        
w\sigma_z \rightarrow {\cal W}[\rho, |{\bf m}|]\,
[m_z\sigma_z + 2(m_+\sigma_- + m_-\sigma_+)]\,.
\label{eq 4}
\end{eqnarray}
In the static case, the inclusion of the densities $m_+$ and $m_-$
makes no difference since they vanish identically. The situation is 
different when the system interacts with a time-dependent field that
couples to the nucleon spin through the general term:
\begin{equation}
{\bf F}\cdot {\bf \sigma}=F_z\sigma_z + 2( F_+\sigma_- + F_- \sigma_+)\,.
\end{equation}  
As a consequence, the interaction Hamiltonian
causing transverse spin excitations may be written as:
\begin{eqnarray}        
H_{\rm int} \sim \sigma_f^-e^{-\imath \omega t}
+\sigma_f^+e^{\imath \omega t}\,.\label{eq 5}
\end{eqnarray}
$H_{\rm int}$ causes non-vanishing variations in the magnetization components $\delta m_+$ and $\delta m_-$ which, in turn, generate at
first-order perturbation theory a variation in the mean field potential. 
Following the steps described in Ref. \cite{Lip16} the TDLSDA response function is given by (once again $V$ is the volume):
\begin{eqnarray}        
\chi_t(q,\omega)=\frac{\chi_t^0(q,\omega)}{ 1-\frac{2}{V}{\cal 
W}(\rho,m)\chi_t^0(q,\omega)}\,,\label{eq 11}
\end{eqnarray}
where $\chi_t^0(q,\omega)$ is the free transverse linear response. 
In the $q v_F\ll\epsilon_F$ limit, where $v_F=k_F/m$ is the Fermi velocity,  it is given by:
\begin{eqnarray}        
\frac{\chi_t^0(q,\omega)}{V}= -\frac{3}{4}\frac{\rho}{\epsilon_F}\left(1+\frac{\omega}{2q v_F}
\ln\frac{\omega-\omega_a-q v_F}{\omega-\omega_a+q v_F}\right)\,,\label{eq 11b}
\end{eqnarray}
where:
$$
\omega_a=\frac{\omega_L}{\left(1+\frac{3\rho{\cal W}(\rho,m)}{2\epsilon_F}\right)}=\frac{2}{3}\frac{k_F^2}{m}\xi~~,
$$
and the last step has been obtained by using relation (\ref{eq 11111}).

The imaginary part of Eq. (\ref{eq 11}) provides the excitations strengths 
$S^{\pm}(q,\omega)=\sum_n|\langle n|\tau_f^{\pm}|0\rangle|^2\delta(\omega-\omega_{no})$ corresponding to the $\Delta S_z=\pm1$ channels,
respectively, through the relation:
\begin{eqnarray}
S^-(q,\omega)-S^+(q,-\omega)=-\frac{1}{\pi} {\rm Im}(\chi_t)~~. 
\label{eq 13a}
\end{eqnarray} 
As we did for Eqs. (\ref{resc_chi}) and (\ref{resc_omega}), 
Eqs. (\ref{eq 11b}) and (\ref{eq 11}) can then be recast in the following way using the adimensional variables $s=\omega/(qv_F)$ and $z=3q/(2k_F\xi)$:
\begin{eqnarray}
\frac{\chi_t^0(q,\omega)}{V\nu}\equiv\frac{\chi_t^0(s,z)}{V\nu}= \Omega_{\pm}(s,z)\,,\label{eq 18}
\end{eqnarray}
with
$$
\begin{array}{c}
\nu=m k_F/\pi^2, \\ 
\\
\Omega_{\pm}(s,z)=-\left(1+\frac{s}{2}\ln\frac{s-1-1/z}{s+1-1/z}\right),
\end{array}
$$
and
\begin{eqnarray}
\frac{\chi_t(q,\omega)}{V\nu}\equiv\frac{\chi_t(s,z)}{V\nu}=\frac{\Omega_{\pm}(s,z)}{ 1-2\nu{\cal
W}(\rho,m)\Omega_{\pm}(s,z)}\,.\label{CHI_TRANS}
\end{eqnarray}

\section{Numerical results} 

\subsection{Response and excitation strengths}

The numerical evaluation of the longitudinal and transverse response functions gives access to information about the neutron dynamics. 
The single particle excitations strengths are computed using Eqs. (\ref{S_LONG}) and (\ref{eq 13a}). On the other hand, the poles of Eqs. (\ref{CHI_LONG}) and (\ref{CHI_TRANS}) are the energies of the collective modes of the system.

\begin{figure}[hbt!]
\centerline{\includegraphics[scale=1.00]{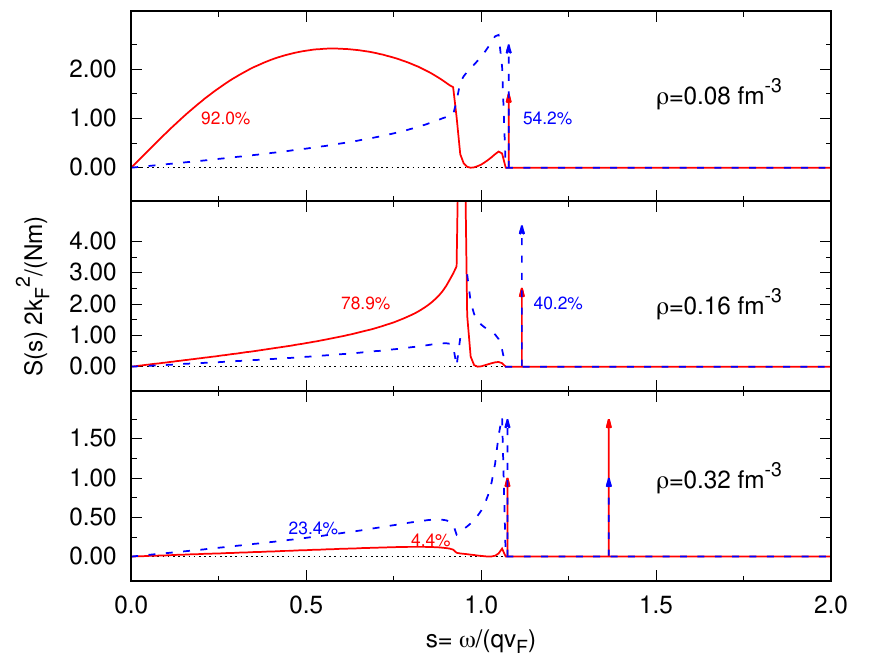}}
\caption[]{Longitudinal Response Function for AV8$^{\prime}$+UIX and spin polarization $\xi=0.2$. The red solid lines and blue dashed lines stand for density and spin density Dynamical Structure Factors (DSFs) respectively. Arrows indicate the presence of a collective mode. The percentages in the plot show the fraction of the total strength pertinent to the particle-hole excitations. The same color scheme holds for Fig.\ref{fig3}-\ref{fig5}}.
\label{fig2}
\end{figure}

\begin{figure}[hbt!]
\centerline{\includegraphics[scale=1.00]{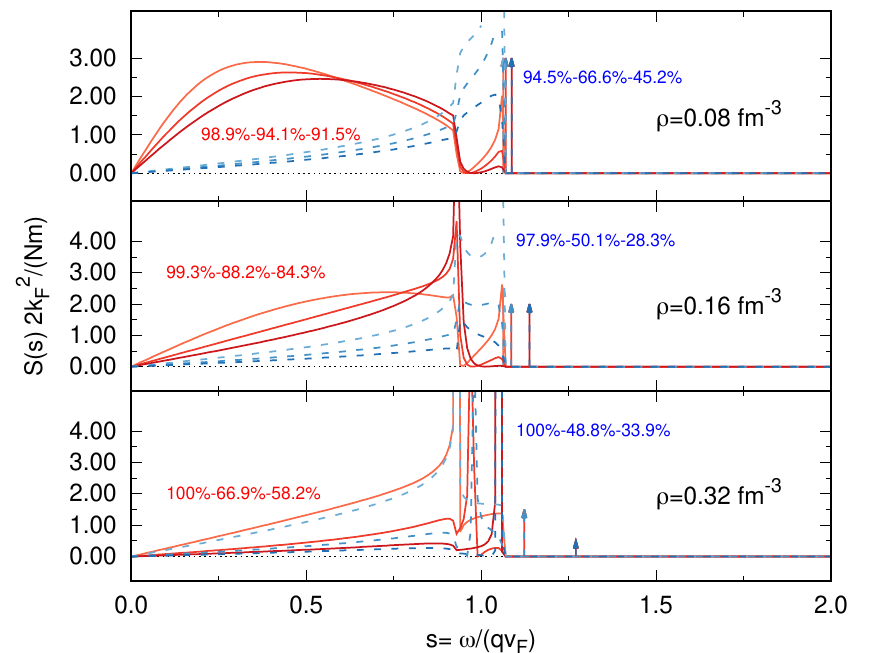}}
\caption[]{Longitudinal Response Function for Chiral Potential at N$^2$LO and spin polarization $\xi=0.2$. In this case we have three lines for each DSF, since we keep track of the errorbars obtained using Chiral effective interaction.}
\label{fig3}
\end{figure}

We report in Fig.\ref{fig2} and Fig.\ref{fig3} the results we obtained for the calculation of the longitudinal responses for the two different potentials used, i.e. the phenomenological AV$8^{\prime}$+UIX interaction and the Local Chiral potential at N$^2$LO. The plots are made as functions of the adimensional quantity $s=\omega/(qv_F)$ for a fixed value of the spin polarization $\xi=0.2$, and for three different values of the density which are characteristic of the outer core of a neutron star ($\rho=0.08,0.16$, and $0.32$ fm$^{-3}$).

\begin{figure}[hbt!]

\centerline{\includegraphics[scale=1.00]{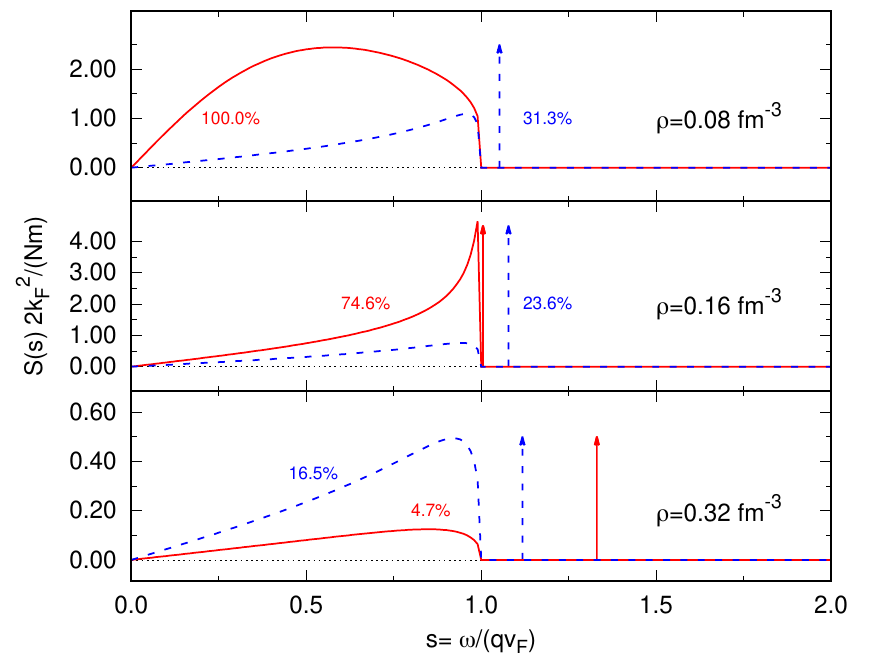}}

\caption[]{Longitudinal Response Function for AV8$^{\prime}$+UIX and spin polarization $\xi=0.0$, i.e. PNM.}
\label{fig4}
\end{figure}

\begin{figure}[hbt!]

\centerline{\includegraphics[scale=1.00]{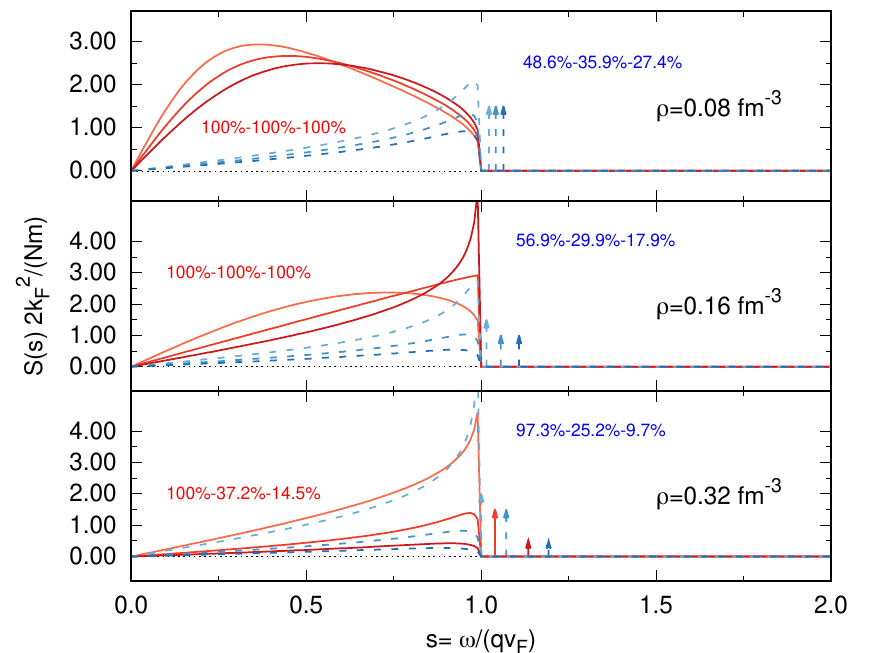}}

\caption[]{Longitudinal Response Function for Chiral Potential at N$^2$LO and spin polarization $\xi=0.0$, i.e. PNM.}
\label{fig5}
\end{figure}

In Fig.\ref{fig4} and Fig.\ref{fig5} the same quantity is reported for spin unpolarized neutron matter. The percentages reported in the graphs show the fraction of the total strength relative to the particle-hole contribution. Arrows represent the presence of collective modes (the size is not proportional to the strength). 
For the response computed using the N$^2$LO potential we propagated the theoretical uncertainty. As expected, at the lowest density considered the results are qualitatively and quantitatively very insensitive to the specific interaction used. At saturation density and above, the theoretical uncertainty on the pressure reflects in a more pronounced difference in the characterization of the single particle spectrum, in particular for as concerns the scalar channel in the region around $\omega=qv_F$. The vector channel is somewhat less affected, at least qualitatively, by the theoretical uncertainty. 
A similar behavior concerns the collective modes. The energy of the collective modes strongly depends on the stiffness of the equation of state. A consequence is that the energy of the collective modes when increasing the density results significantly higher in the AV$8^{\prime}$+UIX case. It should be noticed that in pure neutron matter collective modes are not present for the lowest density considered in the scalar channel. On the contrary, the results at $\xi=0.2$ always show the presence of a collective mode.

\begin{figure}[hbt!]

\centerline{\includegraphics[scale=1.00]{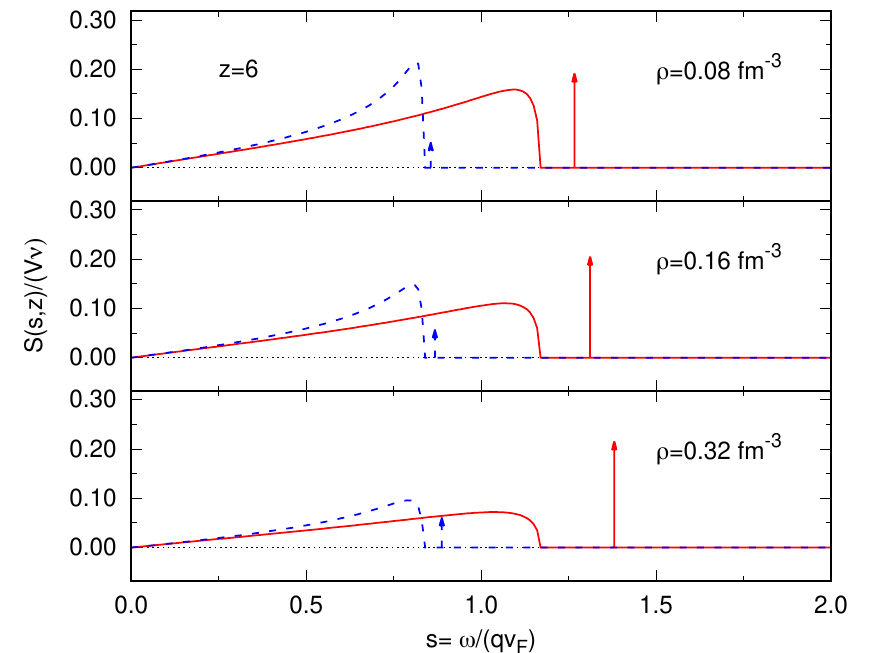}}

\caption[]{Transverse Response Function for AV8$^{\prime}$+UIX at low spin polarization $(z=6)$. Recall that $z=\frac{3q}{2 k_F \xi}$, so $z>1$ means small $\xi$. The full and dashed lines indicate the particle/hole and collective strengths in the $\Delta S_z =-1$ ($s>0$ - red) and $\Delta S_z =+1$ ($s<0$ - blue, which as been plotted flipped and in the $s>0$ region) channels respectively. Same color-scheme holds for Fig. \ref{fig7}.}
\label{fig6}
\end{figure}

\begin{figure}[hbt!]

\centerline{\includegraphics[scale=1.00]{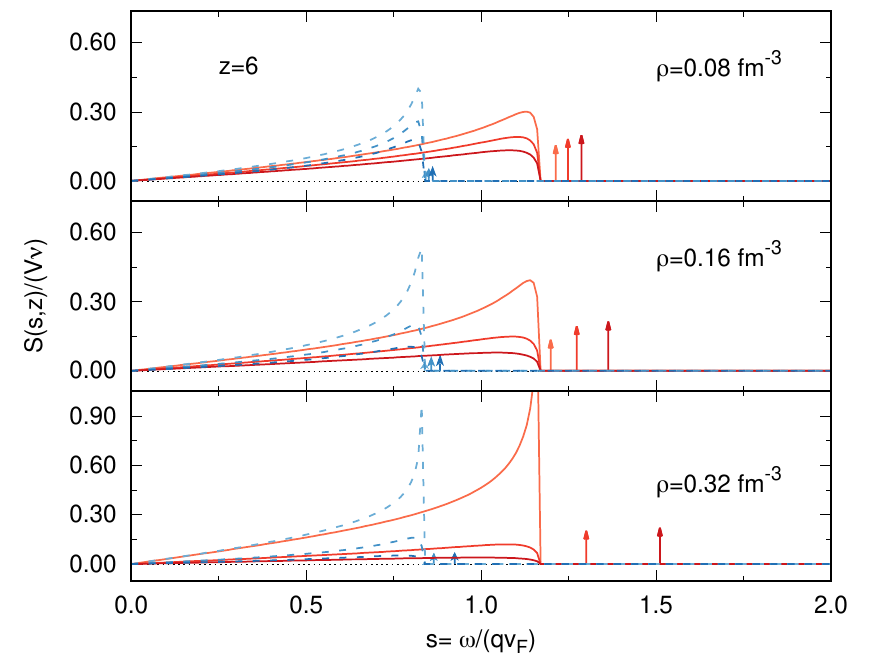}}

\caption[]{Transverse Response Function for Chiral Potential at N$^2$LO and spin polarization $(z=6)$.}
\label{fig7}
\end{figure}

For the transverse response, positive values of $s$ describe the excited states in the $\Delta S_z=-1$ channel, while for negative values of $s$ the excited states  in the $\Delta S_z=+1$ channel.
In Figs. \ref{fig6} and \ref{fig7} we show the results for the transverse response function. In this case, instead of fixing the polarization we fixed the value $z=3q/(2k_F\xi)=6$, still corresponding to a case of low magnetization. The results are qualitatively very close to those obtained for the longitudinal channel, although the dependence on the specific choice of the interaction results weaker, both for the particle-hole and the collective part of the spectrum. 

\subsection{Neutrino mean free path}

The neutrino mean free path (NMFP) can be computed by integrating the total excitation strength $S(q,\omega)$ (in both the longitudinal and transverse channels), to first obtain the total neutrino cross section $\sigma$ \cite{Iwa82,Cow04}:
\begin{widetext}
\begin{eqnarray}
\sigma=\frac{G_F^2}{2}\frac{1}{E}\int d q\int d\omega(E-\omega)q\left[1+\frac{E^2+(E-\omega)^2-q^2}{2E(E-\omega)}\right]S(q,\omega)
\,,\label{eq 199}
\end{eqnarray}
\end{widetext}
where $E$ is the incident neutrino energy, and $G_F=1.166\times10^{-5}$ GeV$^{-2}$.
Integration must be performed on a region of $q$ and $\omega$ compatible with the scattering, as discussed for instance in
Ref. \cite{Iwa82}. We will assume neutrinos to be ultra-relativistic and non-degenerate. 
The NMFP $\lambda$ can be derived from the total neutrino cross section $\sigma$ from the relation $\lambda=1/(\sigma\rho)$.

From existing estimates of neutron spin susceptibility \cite{Fan01}, we expect the induced spin polarization to be low even in presence of strong magnetic fields. 

\begin{figure}[h!]
\centerline{\includegraphics[scale=1.00]{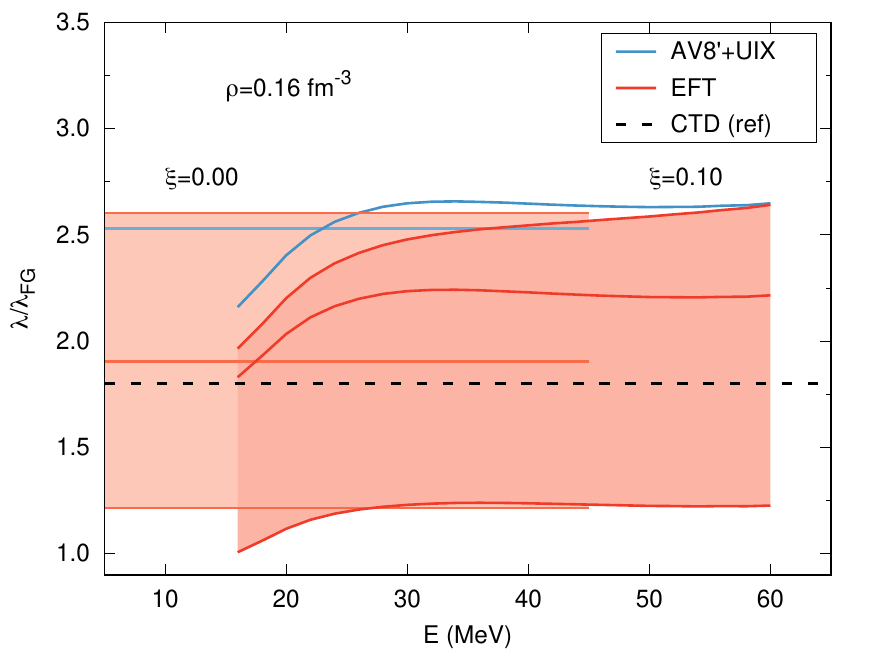}}
\caption[]{Neutrino Mean Free Path ratio with respect to the Free Fermi Gas at saturation density $\rho_0=0.16$ fm$^{-3}$ for spin polarization $\xi=0.1$ and for PNM ($\xi=0$).}
\label{fig8}
\end{figure}

In Fig.\ref{fig8} we report the results we obtained at saturation density for spin polarization $\xi=0.0$ and $\xi=0.1$ and compared them with the result obtained for PNM with a more refined method \cite{Lov14}. The NMFP for spin unpolarized pure neutron matter is essentially independent of the incident energy of the neutrinos. The presence of a small spin-asymmetry shows instead non trivial patterns, significantly increasing the neutron matter opacity for low neutrino energies.

The estimated theoretical uncertainty on the results computed from the chiral interaction are quite significant. Nevertheless, the prediction obtained making use of the phenomenological interaction differ of about 20\% from that of the N$^2$LO potential, close to the upper limit predicted by the propagated uncertainty.
The comparison with previous work done using the Tamm-Dancoff approximation \cite{Lov14} shows that while the NMFP of about $1.8\lambda_{FG}$ is in good agreement with that predicted by the N$^2$LO potential, it is about 30\% lower than that obtained with the Argonne/Urbana potential which represents a more fair comparison.

\begin{figure}[h!]
\centerline{\includegraphics[scale=1.00]{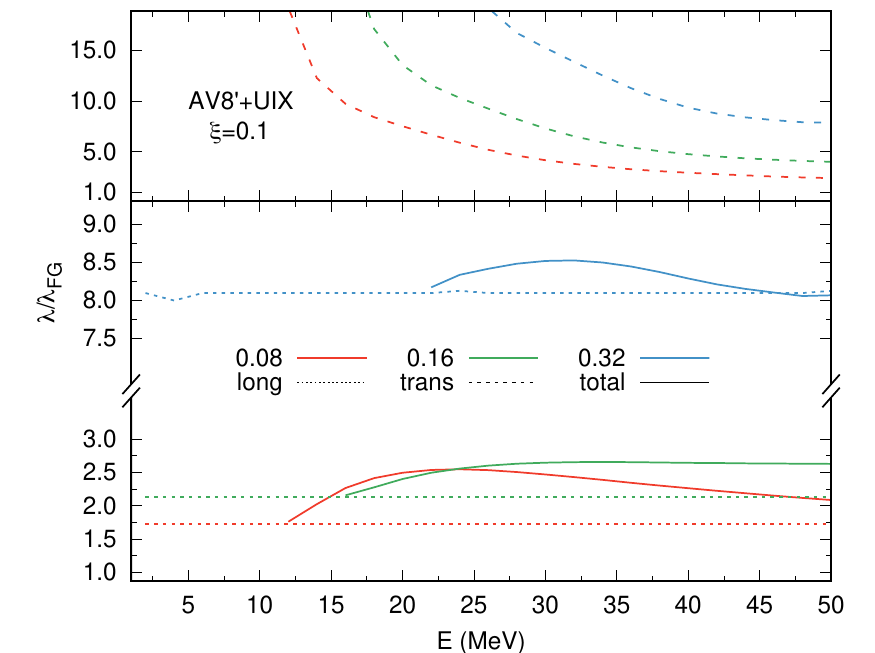}}
\caption[]{Neutrino Mean Free Path ratio with respect to the Free Fermi Gas for spin polarization $\xi=0.1$ as a function of density (color on line). Dotted lines are the contributions coming from the longitudinal channels, dashed lines from the transverse part, while solid lines show the total mean free path.}
\label{fig9}
\end{figure}

In Fig.\ref{fig9} we show the contribution of the different channels to the total neutrino mean free path. As an example we report the results for the phenomenological potential AV$8^{\prime}$+UIX at spin polarization $\xi=0.1$. 

Results are plotted at saturation density $\rho_0=0.16$ fm$^{-3}$, half and twice saturation density. We observe that at all densities the contribution coming from the longitudinal part is almost constant as a function of the energy of the incident neutrino. We observe that for both channels NMFP increases with the density. However, since the relative weight of the two contributions is different for each densities the result gives a total NMFP with non-trivial density dependence. 

\begin{figure}
\centerline{\includegraphics[scale=1.00]{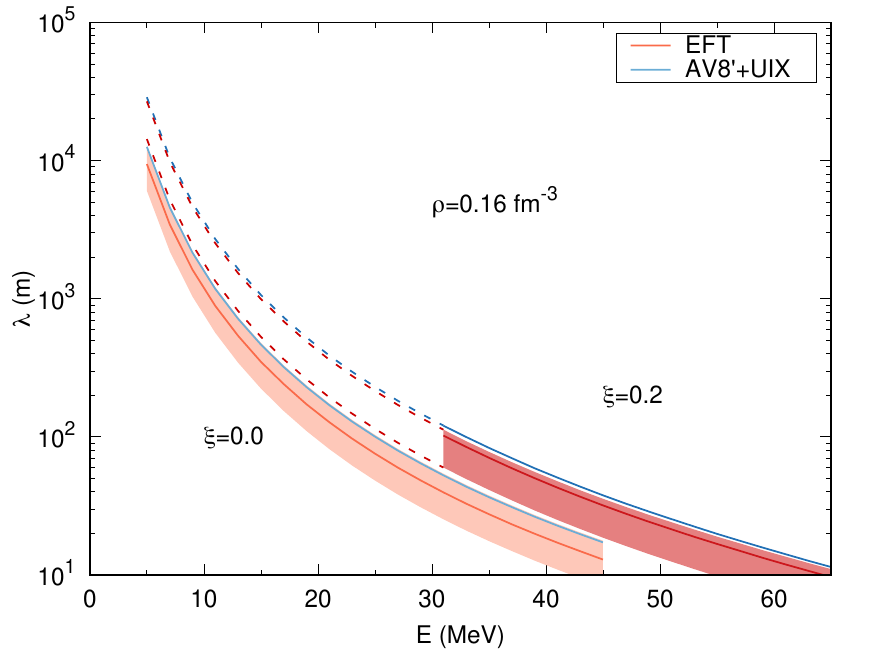}}
\caption[]{Neutrino Mean Free Path for PNM and for spin polarization $\xi=0.2$ as a function of incident neutrino energy. In PNM the longitudinal and transverse channel contribute equally to the total NMFP, while as soon as there is some spin polarization we can observe an energy threshold under which the NMFP is entirely determined by the longitudinal response (dotted lines). The same behavior can be seen also in Fig.\ref{fig9} at various densities.}
\label{fig10}
\end{figure}

To understand the implication of spin-polarization to the NMFP we show in Fig.\ref{fig10} the NMFP in function of the energy of the incident neutrino. The NMFP has to be compared to the radius of the neutron star ($\approx 1.2-1.5\cdot10^4$ m): above this value matter is essentially transparent to neutrinos, while the typical energies of the neutrinos of astrophysical interest are in the range $0.1-50$ MeV \cite{Red99,Eji00}.  

\section*{Conclusions}

We successfully extended TDLDA to study the response function of neutron matter with arbitrary spin polarization both in the longitudinal and in the transverse channel starting from accurate QMC calculations of the equation of state for PNM and for SPPNM. We employed two different neutron-neutron potentials, the phenomenological AV8'+UIX and a modern local chiral EFT potential. For the latter, we considered the predicted theoretical uncertainties coming from the expansion scheme of the theory. We computed estimates for the NMFP showing non trivial contribution coming from the two different channels and also the effects of a small spin polarization, which could play a role in high energy phenomena such as neutron star mergers and supernova explosions. At the neutron core conditions matter is essentially transparent to neutrinos, while relevant effects could be seen in the neutron star crust. 

\section*{Acknowledgments}

We thank Alessandro Lovato and Omar Benhar for useful discussion about the subject of this paper. Calculations were performed partly at CINECA under the INFN supercomputing grant for the MANYBODY collaboration and we also used resources provided by NERSC, which is supported by the US DOE under Contract DE-AC02-05CH11231. Computational resources have been also provided by Los Alamos Open Supercomputing.
The work of S.G. was supported by the NUCLEI SciDAC program,
by the U.S. DOE under contract DE-AC52-06NA25396, by the LANL LDRD
program, and by the DOE Early Career Research Program.


\end{document}